\documentclass[letterpaper,journal]{IEEEtran}
\usepackage[nocompress]{cite}
\usepackage{xcolor}
\usepackage{amsmath,amssymb,amsfonts}
\usepackage{algorithmic}
\usepackage{graphicx}
\usepackage{textcomp}
\usepackage{multirow}
\usepackage[flushleft]{threeparttable}
\usepackage{orcidlink}
\usepackage{hyperref}
\usepackage{stfloats}
\usepackage{url}
\usepackage{verbatim}



\hypersetup{colorlinks = true, urlcolor = blue, linkcolor = blue, citecolor = blue}

\begin{document}
	
	\title{Electromagnetic Investigation of Crosstalk in Bent Microstrip Lines with Partial and Apertured Shielding: Simulations and Measurements}
	
	\author{Mohammad Eskandari\orcidlink{0009-0002-0599-2742},~\IEEEmembership{Student Member,~IEEE}, and Mojtaba Joodaki\orcidlink{0000-0002-2239-1271},~\IEEEmembership{Senior Member,~IEEE}
		\thanks{Mohammad Eskandari and Mojtaba Joodaki are with the School of Computer Science and Engineering, Constructor University, Campus Ring 1, 28759 Bremen, Germany (e-mail: meskandari@constructor.university; mjoodaki@constructor.university).}}
	
	\maketitle
	
	\begin{abstract}
		This paper presents an electromagnetic investigation of the crosstalk between two bent microstrip lines (MLs) separated by a perforated planar shield. We present the first combined simulation and experimental study investigating the impact of structural discontinuities in both the microstrip lines and the shield on crosstalk, along with a detailed analysis of the underlying electromagnetic mechanisms. The discontinuities analyzed include bends in the MLs, variations in the shield height, partial shielding, and apertures in the shield. Furthermore, multimodal wave theory in a rectangular waveguide is applied to predict crosstalk behavior when the shield contains an aperture. This study aims to conceptually elucidate complex crosstalk phenomena that are difficult to model using circuit theory. The predicted crosstalk behavior for different practical cases shows strong agreement with both simulations and measurements, demonstrating the accuracy of the modeling and the experimental setup.
	\end{abstract}
	
	
	\section{Introduction}
	\IEEEPARstart{C}{rosstalk}, even when appearing as minor signal disturbances, can compromise the reliability of critical applications such as microwave sensing, medical instrumentation, and aerospace systems. In high-volume manufacturing, unresolved crosstalk issues discovered in later stages can lead to costly redesigns and failures in compatibility testing \cite{b1}. This problem is further amplified in high-speed electronic systems that operate at higher frequencies and are increasingly compact within multilayered configurations. Consequently, the parasitic coupling between densely packed traces can severely degrade signal integrity, distort timing margins, and limit system bandwidth \cite{c0,c1,c2,c3}. Moreover, most modern devices integrate RF, digital, and analog subsystems within compact footprints \cite{c6}, and superconducting design \cite{c7}, making microstrip crosstalk a significant factor affecting device compatibility \cite{c5}.
	
	The concept of crosstalk has been studied since the early days of electrical and electronic engineering. In 1934, Schelkunoff proposed an electromagnetic theory for coaxial transmission lines (TLs) and cylindrical shields \cite{b2}. Later, he explored the crosstalk implications of this theory based on distributed mutual impedance and introduced the concepts of near-end crosstalk (NEXT) and far-end crosstalk (FEXT) \cite{b3}. Since then, numerous studies have addressed the calculation of crosstalk between TLs \cite{b4,b5,b6,b7,c8}. Additionally, various techniques have been proposed to reduce NEXT, FEXT, or both. These include inserting a guard trace (typically grounded with vias) \cite{b8,b9,b10,b11}, increasing spacing between TLs \cite{b12,b13}, periodically corrugating the TLs \cite{b14,b15,c4} or the guard trace \cite{b16}, etching the substrate \cite{b17}, using a superstrate \cite{b18}, and employing decoupling capacitors \cite{b19}. Since crosstalk originates from the radiated fields of a microstrip line (ML), many studies have also analyzed and characterized different types of radiation from MLs \cite{b20,b21,b22,b23}.
	
	\begin{figure*}[t]
		\centering
		\includegraphics[width= 6.5in]{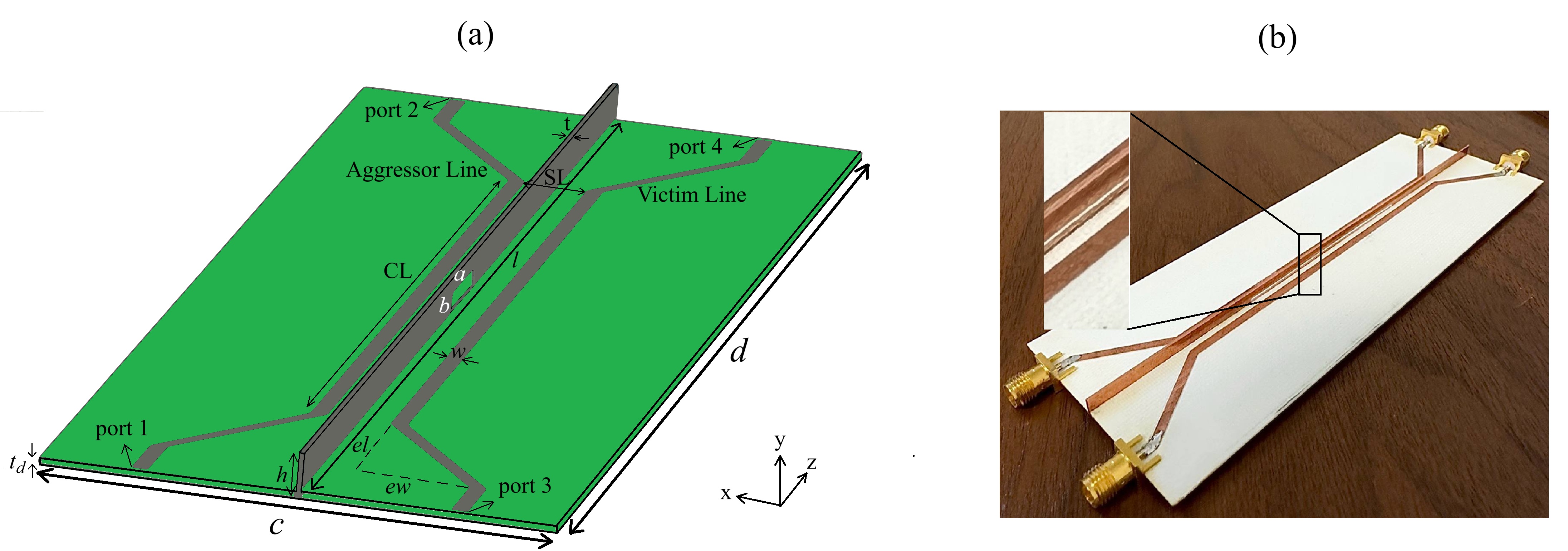}
		\caption{(a) Geometry and (b) a sample of fabricated PCBs of two bent MLs implemented on a Rogers RO4003C substrate, and symmetrically separated by a copper made shield with an aperture (with the height of 1 mm).}
		\label{fig1}
	\end{figure*}
	
	The isolation of circuit blocks is essential in electronic systems to reduce interference between functional units. One widely used isolation technique is circuit partitioning, which physically and electrically separates components according to their operational roles. A fundamental approach involves the use of planar conductive barriers to suppress unintentional coupling or crosstalk. However, practical design constraints such as interconnects, ventilation requirements, and limited space allow some leakage of electromagnetic energy. This leakage can significantly degrade signal integrity by increasing crosstalk between transmission lines serving sensitive circuit components. To mitigate these effects, designers typically adjust the routing and spacing of transmission lines relative to shielding structures. Despite the widespread use of circuit partitioning and shielding, comprehensive studies on the impact of partial or imperfect shielding on crosstalk remain limited.
	
	This paper extends our previous investigation in \cite{b24} by experimentally measuring and analyzing, for the first time, the influence of imperfect shielding on crosstalk between two MLs. To emulate a realistic routing scenario, and following approaches in prior works such as \cite{b12,b13}, the MLs are bent so that they are closest to each other over a designated section, referred to as the coupling length (CL). This study represents the first crosstalk measurement of partial and apertured shielding between two MLs. The effects of this bending, together with imperfect planar shielding, are investigated through 3D full-wave simulations using CST Studio Suite 2023\textregistered, and validated through measurements performed with a Rohde \& Schwarz ZVB20 Vector Network Analyzer (VNA), as explained in Section~\ref{sec2}.

	In Section~\ref{sec3}, first, we analyze the impact of bent MLs, which exhibit strong coupling within their CL. Next, the effect of partial shield coverage is examined. After that, the impact of shield height on crosstalk is evaluated, and the limitations of our previously proposed approximate crosstalk expression are discussed. Finally, the impact of an aperture and a gap in the shield on crosstalk is investigated, providing a different perspective on our previous efforts regarding apertures in conducting shields in the electromagnetic compatibility (EMC) scope \cite{bo1,bo2,bo3}. Our approach is based on electromagnetic wave theory rather than conventional circuit modeling. The geometrical theory of diffraction (GTD) \cite{b25} is employed to predict the effects of shield diffraction on crosstalk. Furthermore, our recent electromagnetic analyses in this area \cite{b26,b27} allow us to treat an aperture in the shield as a waveguide, enabling an electromagnetic exploration of its impact on crosstalk behavior.

	\section{Geometry and Measurement Setup}
	\label{sec2}
	Fig.~\ref{fig1}(a) illustrates the geometry of two identical bent MLs separated by a grounded planar shield. The structure was implemented on a Rogers RO4003C printed circuit board (PCB) substrate with a relative permittivity of $\varepsilon_r = 3.38$ and a loss tangent of $\tan\delta = 0.0027$ at 10~GHz. To achieve a characteristic impedance of 50~$\Omega$, the physical dimensions of the parameters shown in Fig.~\ref{fig1}(a) were calculated for PCB fabrication and are listed in Table~\ref{tab1}. Fig.~\ref{fig1}(b) shows the fabricated PCB designed by the given parameters including an aperture in the shield.
	
\begin{figure}[!t]
		\centering
		\includegraphics[width= 2.7 in]{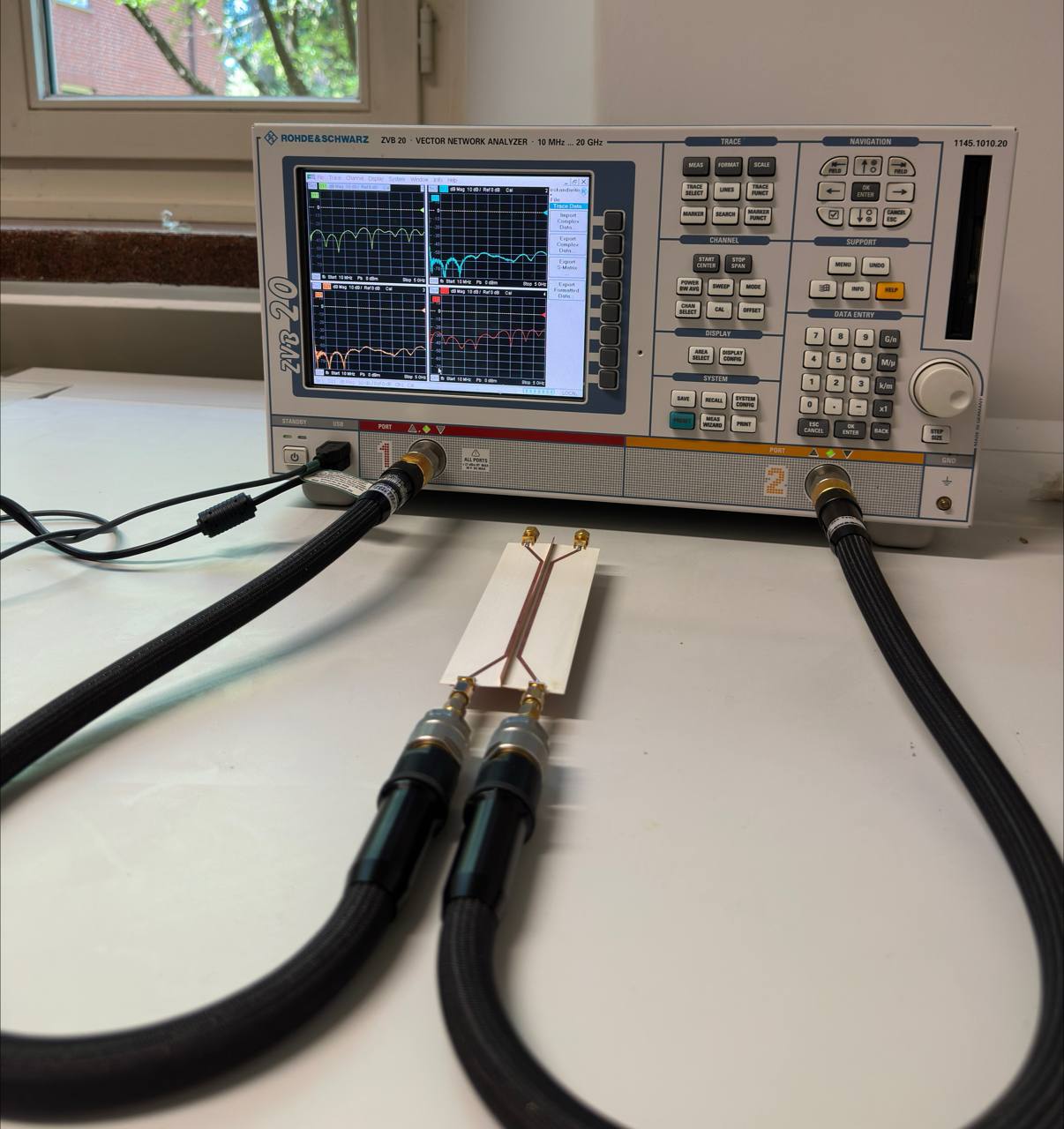}
		\caption{The experimental setup used for the crosstalk measurement while the VNA is measuring two-ports scattering parameters. This configuration is set for NEXT measurement of the fabricated PCB with a solid copper made shield while the other two SMA ports are terminated with 50~\(\Omega\) standard loads.}
		\label{fig2}
	\end{figure}

	For the experimental measurements, the two-port scattering parameters were measured using a Rohde \& Schwarz ZVB20 VNA, rated for measurement frequencies up to 20~GHz and calibrated for a 50~$\Omega$ reference impedance using a standard short--open--load--through (SOLT) procedure. The VNA was connected to the PCB through SMA-50-0-9/111\_NE female connectors, rated for operation up to 18~GHz. Parasitic effects from connectors and cables were minimized through careful calibration and by terminating the remaining two ports with standard 50~$\Omega$ loads. The measurement setup is shown in Fig.~\ref{fig2}. 
    	
	Most of the measurement results are presented for frequencies up to 5~GHz, while selected cases extend to 6~GHz to more clearly illustrate the resonance behavior. To realize all required test structures, only two boards were fabricated: one with the substrate cut in the middle to accommodate different shield placements, and one uncut. This approach enables measurements of all practical shield configurations, including those with apertures, while minimizing errors caused by fabrication tolerances between different PCBs.

	\begin{table}[t]
		\caption{Dimensional Parameters of the Proposed Geometry}
		\begin{center}
			\begin{tabular}{|c|c|c|}
				\hline
				\textbf{\textit{Parameter}}& \textbf{\textit{Value}}& \textbf{\textit{Description}} \\
				\hline
				$a$ & 60 mm & Aperture Length \\
				\hline
				$b$ & 1 mm & Aperture Height \\
				\hline
				$c$ & 50 mm & Ground Width \\
				\hline
				$d$ & 150 mm & Ground Length \\
				\hline
				$CL$ & 100 mm & Coupling Length \\
				\hline
				$el$ & 20 mm & Line Extension in Length \\
				\hline
				$ew$ & 11.5 mm & Line Extension in Width \\
				\hline
				$h$ & 4.8 mm & Shield Height \\
				\hline
				$l$ & 150 mm & Shield Length \\
				\hline
				$SL$ & 3.6 mm & Separation Length \\
				\hline
				$t_d$ & 0.813 mm & Dielectric Thickness \\
				\hline
				$t$ & 0.5 mm & Shield Thickness \\
				\hline
				$w$ & 1.8 mm & Line Width \\
				\hline
				$xt$ & 35 $\mu$m & Traces' Thickness \\
				\hline
			\end{tabular}
			\label{tab1}
		\end{center}
	\end{table}

	\section{Results and Discussion}
    \label{sec3}
	\subsection{Bends and Their Impacts}
	\label{sec3a}
	\begin{figure*}[!t]
		\centering
		\includegraphics[width= 5.8 in]{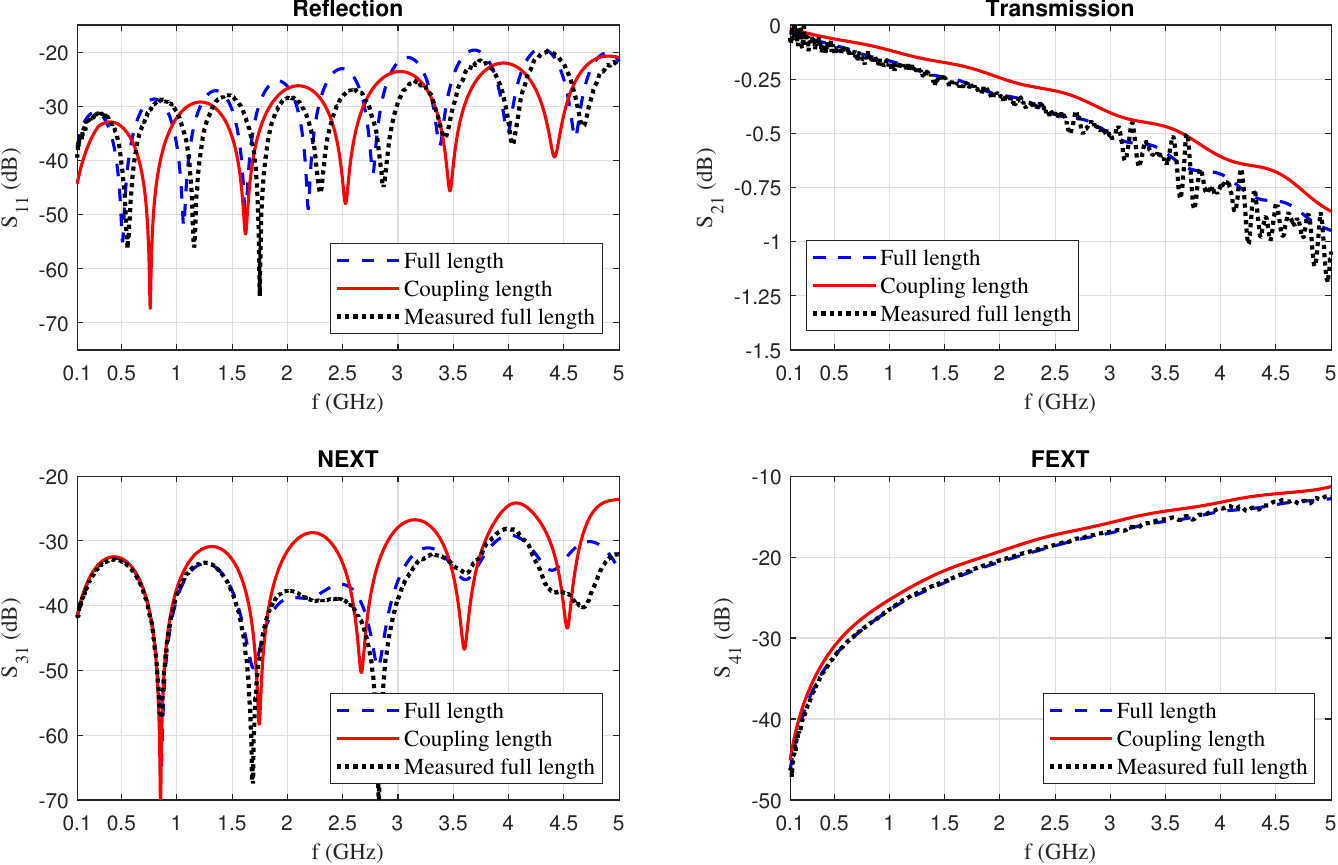}
		\caption{Scattering parameters of the proposed geometry without the shield, modeled using the dimensional parameters in Table~\ref{tab1}. In the first case, the MLs are terminated at the port locations indicated in Fig.~\ref{fig1}, while in the second case, they are terminated at both ends of the CL on both lines. The measured results correspond to the first case.}
		\label{fig3}
	\end{figure*}
    In our previous research \cite{b24}, we theoretically discussed the crosstalk between two MLs along a CL without any bends. However, in practical PCB layouts, most traces are bent to accommodate integration and connection requirements. These bends can alter the ML characteristics, which in turn affect the resulting scattering parameters, as generally discussed in reference books such as \cite{b28}. To investigate the impact of bends in our study, we compare the unshielded case of Fig.~\ref{fig1}(a) with the unshielded case studied in \cite{b24}, where the dimensional parameters are the same as listed in Table~\ref{tab1}. The simulated scattering parameters, including the measured results for the first case, are shown in Fig.~\ref{fig3}.
	
	The ML impedance varies with frequency due to the dispersion-dependent effective permittivity, which can be modeled using the Hammerstad–Jensen approach \cite{b30}. Thus, even for a matched design as presented here, minor differences between the line and port impedances occur over frequency, causing $S_{11}$ of a finite-length transmission line to exhibit sinusoidal ripples (cancellation behavior) due to the load effect on the reflection coefficient \cite{b28}. For a ML, these cancellation frequencies can be approximated by
	\begin{equation}
		f_x \backsimeq \frac{c_0 N}{2 l_e \sqrt{\epsilon_{\text{r,eff}}}}
	\end{equation}
	where $c_0$ is the speed of light in free space, $l_e$ is the electrical length of the ML, $\epsilon_{\text{r,eff}}$ is the effective relative permittivity of the ML, and $N = 1,2,3,...$.
	
	The cancellation frequencies for the first and second cases are $0.59N$~GHz and $0.92N$~GHz, respectively, which are reasonably aligned with the results in Fig.~\ref{fig3}. As discussed in \cite{b24}, in addition to impedance differences, the effect of the adjacent line must be included by increasing the effective relative permittivity to obtain accurate cancellation frequencies. Within the coupling length, the equivalent ML characteristics change, leading to similar impedance differences at both ends of the CL. Hence, a similar cancellation behavior is expected for the NEXT values, with two virtual source ports (used for $l_e$) positioned at both ends of the coupling length on the victim line. A comparable formula for the maxima of the NEXT values, based on bi-directional coupled-mode theory \cite{b6}, is presented in \cite{b24}, providing a similar electromagnetic interpretation of this phenomenon.
	
	As shown in Fig.~\ref{fig3}, the cancellation behavior for $S_{11}$ and NEXT values in the first case weakens beyond $f=2$~GHz. This is attributed to the bent parts of the MLs, whose mismatches increase with frequency. Since the length of the bent parts is approximately 23~mm, (1) predicts that the first maximum from the bends occurs around half of their cancellation frequency, i.e., roughly 2~GHz. Additionally, mismatches differ at both ends of the oblique parts due to the presence of the adjacent line, further weakening cancellations beyond this frequency.
    
	\begin{figure*}[!t]
		\centering
		\includegraphics[width= 6 in]{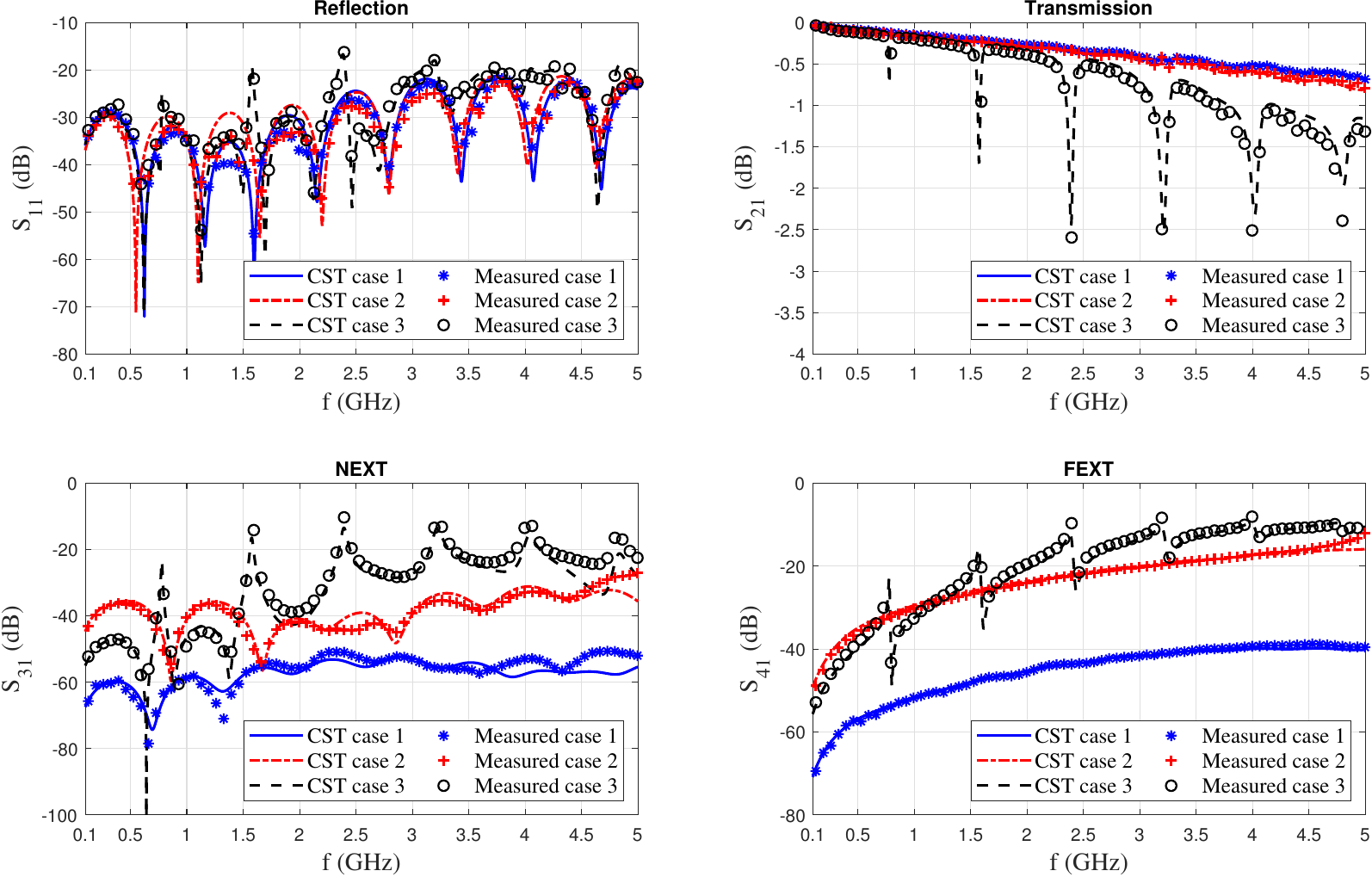}
		\caption{Scattering parameters of the proposed geometry under three shielding configurations without the aperture. Case 1: full shield implemented as specified in Table~\ref{tab1}. Case 2: shield covering only the dielectric region. Case 3: shield covering only the air region, with the same height as in Case 1 and grounded at both ends.}
		\label{fig4}
	\end{figure*}
	As previously noted, a bent ML exhibits different characteristics compared to a straight ML. These differences primarily increase the average reflected power to the source ($S_{11}$), thereby reducing the average transmitted ($S_{21}$) and coupled (NEXT and FEXT) powers, as shown in Fig.~\ref{fig3}. Considering the impact of bends on crosstalk, it can be concluded that the main coupling occurs within the closest region of the MLs (i.e., the CL), while the additional bent traces mainly contribute an attenuating effect.

	\subsection{Different Coverage by the Shield}
    \label{sec3b}
	In all shielding applications, the goal is to achieve maximum isolation between sub-circuits. This requires maximizing shield coverage while ensuring a seamless ground connection, often implemented using via-pitch techniques. To this end, the three shield coverage cases discussed in \cite{b24} are now investigated practically using our proposed geometry, with corresponding measurement results shown in Fig.~\ref{fig4}.
	
	In case~3, the presence of a dielectric-filled short-depth waveguide—modeling the shield when the grounded shield acts as a rectangular metal frame enclosing the dielectric—can generate significant fields near its apertures. These fields are dominant around depth-dependent cut-off frequencies of the waveguide, and can sharply alter the magnitudes of all scattering parameters, as shown in Fig.~\ref{fig4}. These frequencies are approximately $0.79N$~GHz, which are greater than the actual (i.e., infinite-length) $TE_{N0}$ cut-off frequencies of the waveguide as
	\begin{equation}
		f_c = \frac{c_0 N}{2 d \sqrt{\Re e(\epsilon_\text{r})}}
	\end{equation}
	yielding $0.55N$~GHz for the given parameters.
	
	Excluding the results of case~3, the presence of the shield increases $S_{21}$ compared to Fig.~\ref{fig4} by blocking direct coupling paths and reflecting a portion of the radiated wave back to the line. The coupling path in these two cases (i.e., cases~1 and 2)  passes primarily over the shield edges. Increasing the shield height extends the coupling path distance, reducing the significance of coupled fields along the CL and making them comparable to the coupled fields outside the CL (i.e., the bent parts). As shown in Fig.~\ref{fig4}, while the cancellation frequencies for the NEXT values of the dielectric shield largely follow the behavior observed in the absence of a shield, the cancellation frequencies for the NEXT of the complete shield decrease and tend to align with (1), similar to $S_{11}$.
	
	The results show that partial shielding (cases~2 and 3) does not significantly reduce crosstalk compared to the complete shield. However, when the shield covers the dielectric region, direct coupling through the substrate is blocked, and the remaining coupling is due to fields diffracted from the top edges of the shield that are tangent to the substrate surface. So, instead of increasing the height, it is electromagnetically possible to reduce these diffracted fields by locally modifying the shield geometry (e.g., using specific routing patterns) and further suppress crosstalk such as what presented in \cite{b16}.

	\subsection{Different Shield Height}
	\label{sec3c}
	In \cite{b24}, we proposed an expression based on the GTD to predict crosstalk behavior when the shield height changes. We emphasized that this expression is suitable for behavior prediction rather than exact value estimation for two main reasons: first, the far-field incident assumption becomes inaccurate for short shields; second, it considers only cross-sectional coupling and ignores the length dependency of crosstalk, which is the main factor causing the difference between NEXT and FEXT.
	\begin{figure}[!t]
		\centering
		\includegraphics[width= 3.2 in]{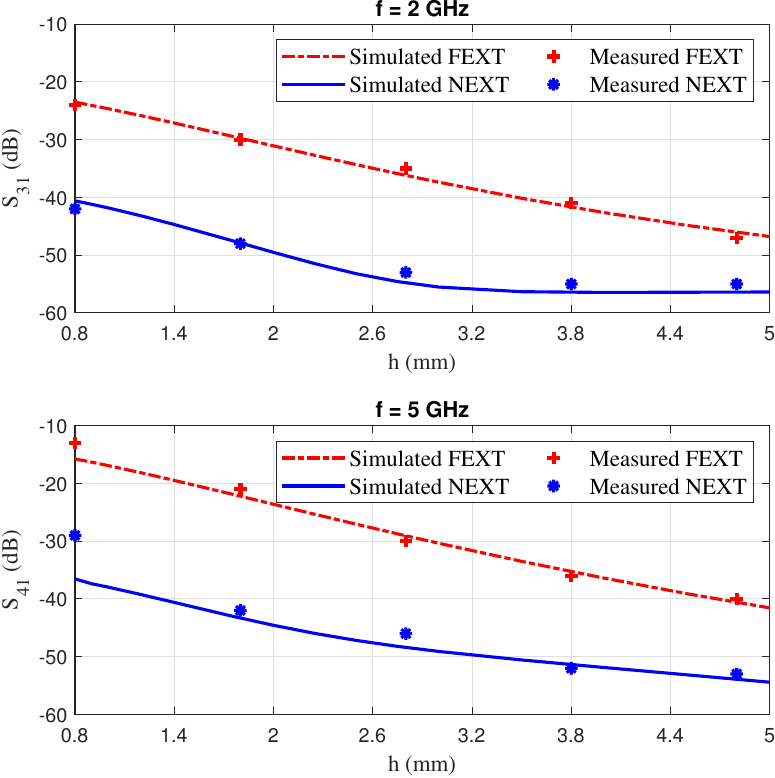}
		\caption{Simulated and measured crosstalk of the proposed geometry in Fig.~\ref{fig1}(a) without an aperture, as a function of shield height $h$, at $f = 2$~GHz and $f = 5$~GHz.}
		\label{fig5}
	\end{figure}
	
	To investigate this effect practically, the simulated and measured crosstalk results for the proposed geometry are shown in Fig.~\ref{fig5}. For the measurements, five shields of different heights were used on the same PCB to minimize fabrication errors. Both simulation and measurement results are in good agreement, indicating that the reduction slope with increasing shield height is smaller for NEXT values than for FEXT values. 
	
	In addition to the effects of bent parts discussed in Section~\ref{sec3b}, this behavior can be explained electromagnetically: increasing the shield height extends the wave traveling path, and at a sufficiently long path (i.e., in the far-field region), a similar field couples to both the near-end and far-end ports, resulting in comparable values for NEXT and FEXT. Since the NEXT values are generally smaller than FEXT, their reduction slope with increasing shield height is correspondingly lower, consistent with this electromagnetic interpretation.

	\subsection{Aperture Effect}
	
\begin{figure}[!t]
		\centering
		\includegraphics[width= 3.3 in]{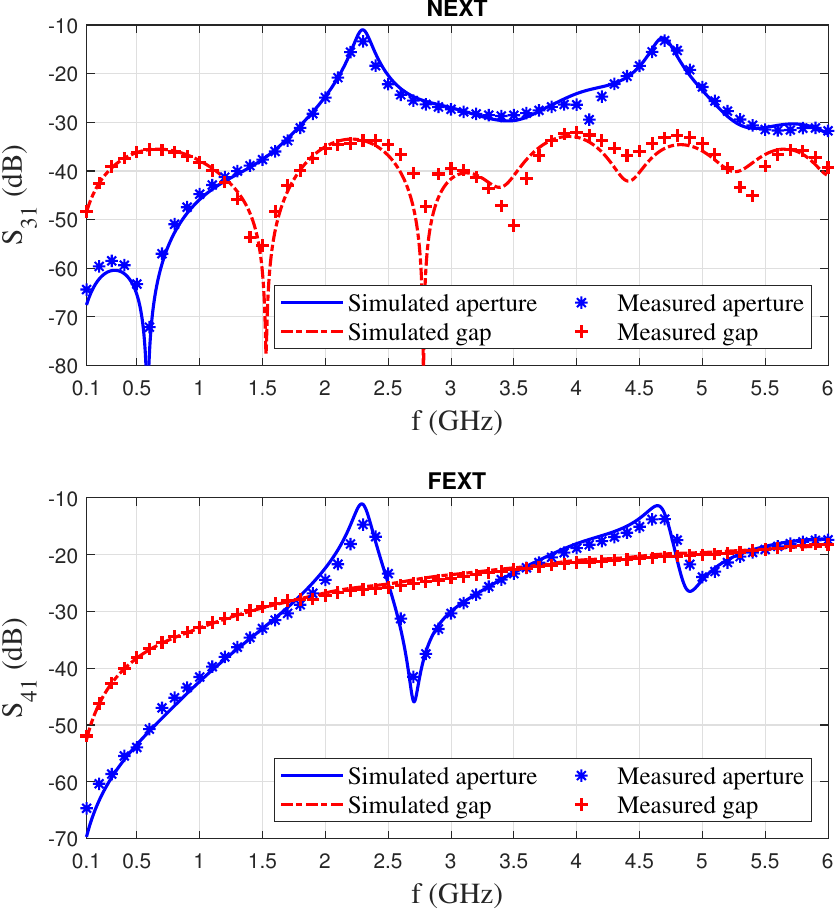}
		\caption{Simulated and measured crosstalk of the proposed geometry for two cases. In the first case, an aperture with dimensions \(a = 60\)~mm and \(b = 1\)~mm is located at the center of the shield, with its bottom edge tangent to the substrate. In the second case, a gap with the same horizontal length (\(a = 60\)~mm) is created above the substrate, resulting in an opening height of \(b = h - t_d \approx 4\)~mm.}
		\label{fig6}
	\end{figure}

	To investigate the impact of an aperture in the shield, we simulated and measured the proposed geometry shown in Fig.~\ref{fig6} for two cases. First, an aperture with its bottom edge tangentially aligned to the substrate; second, a gap created by removing the metal above the aperture of the first case. Since the result of case~1 in Fig.~\ref{fig4} shows strong crosstalk rejection for the complete shield, from an electromagnetic perspective, this gap length can be interpreted as a new equivalent coupling length between the two MLs for the shielded dielectric case (i.e., case~2 in Fig.~\ref{fig4}). To examine this, based on the previously discussed similarity of NEXT cancellation frequencies with (1), $l_e$ can be considered as this equivalent coupling length. Therefore, the expected NEXT cancellation frequencies occur at $1.53N$~GHz. However, only the first frequency aligns with the results in Fig.~\ref{fig6}. This discrepancy is not in conflict with the concept of equivalent coupling length but is due to the impact of the bent parts starting around 2~GHz, as discussed in Section~\ref{sec3a}.
	
	For the aperture case in Fig.~\ref{fig6}, a similar scenario was investigated in Fig.~\ref{fig4}, where the resonance frequencies, given in (2), were derived for an aperture within the dielectric region. The resonance frequencies for a general rectangular aperture, as shown in Fig.~\ref{fig1}(a), are given by \cite{b26}:
	\begin{equation}
		f_c = \frac{c_0}{2\pi} \sqrt{\left(\frac{m\pi}{a}\right)^2 + \left(\frac{n\pi}{b}\right)^2}
	\end{equation}
	where $m$ and $n$ are the mode indices. For the ultra-narrow aperture used in this study, only the TE$_{N0}$ modes are excited within our frequency range, similar to (2). These resonances occur at $2.5N$~GHz, consistent with the results in Fig.~\ref{fig6}. Furthermore, the results show that outside the vicinity of these resonances, particularly at higher frequencies, the crosstalk values of an ultra-narrow aperture are similar to those of a large gap with the same length. This supports the engineering design guideline that noise leakage is minimized by reducing the maximum length of an aperture or crack, rather than minimizing the total surface area without changing the maximum length.
	
	\section{Conclusion}
	
	Any discontinuity, whether in the routing of the MLs or in the coupling path between them, can have a significant effect on the resulting crosstalk, which may be difficult or even impossible to model using conventional circuit theory. Therefore, we investigated crosstalk from an electromagnetic perspective using well-known electromagnetic theories such as the GTD and modal waveguide analysis to predict its behavior. 
	
	It is shown that within the coupling length, where the two MLs are closest to each other, the local characteristics of both MLs change, introducing mismatches at the connections to the rest of the lines. These regions can be modeled by defining two virtual ports at the connection points. These discontinuities, together with the bent portions of the MLs, give rise to complex crosstalk behavior starting at the frequency where the bends begin to influence the fields. 
	
	The shield further modifies the relative contribution of each part to the crosstalk depending on its height, resulting in similar NEXT and FEXT values for sufficiently tall shields. Moreover, as predicted by multimodal waveguide theory, the presence of an aperture in the shield can significantly affect crosstalk, particularly at its mode resonances. At higher frequencies, where the impact of higher-order resonances diminishes, crosstalk is predominantly controlled by the maximum length of the aperture rather than its surface.

	\section*{Acknowledgments}
	
	The authors gratefully acknowledge the technical support provided by Mr. Uwe Pagel at Constructor University, Bremen, Germany.

	\end{document}